%% file: main.tex
\documentclass{article} 
\usepackage{iclr2021_conference,times}

\input{math_commands.tex}

\usepackage{hyperref}
\usepackage{url}
\usepackage{graphicx}
\usepackage{multirow}
\usepackage{widetable}
\usepackage{float}
\newcommand{\email}[1]{\href{mailto:#1}{\texttt{#1}}}

\title{TenSEAL: A Library for Encrypted Tensor Operations Using Homomorphic Encryption}

\iclrfinalcopy

\author{Ayoub Benaissa \\
École Supérieure en Informatique, Sidi Bel Abbès / OpenMined\\
Algeria \\
\texttt{\email{a.benaissa@esi-sba.dz}} \\
\AND
Bilal Retiat \\
École Supérieure en Informatique, Sidi Bel Abbès / OpenMined\\
Algeria \\
\texttt{\email{b.retiat@esi-sba.dz}} \\
\AND
Bogdan Cebere \\
OpenMined \\
\texttt{\email{bogdan.cebere@gmail.com}} \\
\AND
Alaa Eddine Belfedhal\\
École Supérieure en Informatique, Sidi Bel Abbès\\
Algeria \\
\texttt{\email{a.belfedhal@esi-sba.dz}}
}

\begin{document}

\maketitle

\begin{abstract}
Machine learning algorithms have achieved remarkable results and are widely applied in a variety of domains. These algorithms often rely on sensitive and private data such as medical and financial records. Therefore, it is vital to draw further attention regarding privacy threats and corresponding defensive techniques applied to machine learning models. In this paper, we present TenSEAL, an open-source library for Privacy-Preserving Machine Learning using Homomorphic Encryption that can be easily integrated within popular machine learning frameworks. We benchmark our implementation using MNIST and show that an encrypted convolutional neural network can be evaluated in less than a second, using less than half a megabyte of communication.
\end{abstract}

\section{Introduction}
In recent years, we have witnessed the evolution of machine learning as a service (MLaaS). In a typical scenario, the users need to send their input to the service provider, which will execute some algorithms on the data and send back the result. 

This new way of making inferences has two critical issues. First, the users may not want to send their data to the service provider due to privacy concerns. Second, if we do not send users' data to the service provider, we cannot give the users the model due to intellectual property concerns. Using homomorphic encryption, we can follow the same method, except that users' data will always be encrypted. This way, neither the input nor the output will be visible to the service provider, and the evaluation can still happen on this encrypted data.

However, the adoption of homomorphic encryption in machine learning is slow. One reason is that while the available libraries provide an excellent API for cryptographers, they might be challenging to use for data scientists. The other blocker is also the cost for evaluation, both in terms of communication and computation.



\subsection{Contributions}

\begin{itemize}
    \item We present a flexible open-source library for doing encrypted tensor computation using homomorphic encryption. The library can directly convert tensors from popular machine learning frameworks (like PyTorch or Tensorflow) to their encrypted versions.
    
    \item We evaluate a convolution neural network on encrypted data in less than a second, with less than half a megabyte of communication during inference.
\end{itemize}

For the rest of the paper, we describe the library's architecture in Section \ref{sec:library}. Then, in Section \ref{sec:method}, we detail the algorithms needed for evaluating a convolutional neural network in the encrypted space.
In Section \ref{sec:evaluation}, we provide an experimental evaluation of our library, and conclude with some limitations of our work in section \ref{sec:limitations}.

\section{Architecture}
\label{sec:library}

\cite{tenseal} is a library that bridges classical machine learning frameworks to homomorphic encryption capabilities. It manages all the complexities of implementing tensor operations on encrypted data. TenSEAL relies on the implementation of the CKKS (\citet{cheon2017homomorphic}) scheme in \cite{sealcrypto}. The clients can work with plain or encrypted tensors using one of the supported frontend languages (C++ or Python). In a client-server scenario, the message exchange is done using \citet{protobuf}. The core API is built around three main components: the context, the plain tensors, and the encrypted tensors.

\subsection{The TenSEAL context}
The TenSEAL context is the central component of the library. It generates and stores the necessary keys required by an encrypted computation. The context generates the secret-key used for decryption, the public-key used for encryption, the Galois-keys used for rotation, and the relinearization-keys used for relinearization of ciphertexts. This same object will also handle the thread-pool, which controls how many jobs should be run in parallel when performing parallelizable operations. The context can also be configured to do automatic ciphertext relinearization and rescaling during computation.

\subsection{The PlainTensor}
The PlainTensor is a class that connects unencrypted tensors to the encrypted implementations. Figure~\ref{fig:encryptedTensorFlow} in the appendix describes the process of converting the tensors.


\subsection{Encrypted tensors}

The EncryptedTensor interface offers an API that needs to be implemented by every tensor exposed by the library. The interface has a TenSEALContext object, necessary to make any homomorphic encryption operation. The derived classes expose different tensor flavors, such as: 
\begin{itemize}
    \item CKKSVector derives the EncryptedVector interface and can hold a vector of real values by encrypting them into a single ciphertext.
    \item CKKSTensor follows the same strategy as \cite{boemer2019ngraph} and holds an N-dimensional tensor of real values by encrypting them into N-dimensional tensor of ciphertexts. However, it can batch an axis along with the slots available in every ciphertext, thus requiring only an (N-1)-dimensional tensor of ciphertexts.
\end{itemize}

Figures \ref{fig:encryptedTensorConstruction} and \ref{fig:encryptedTensorFlow} in the appendix describe how an encrypted tensor is constructed. From this point onward, we will focus on CKKSVector, as it will be the type used to evaluate the MNIST dataset.

\section{Method}
\label{sec:method}

When building a tensor over a Homomorphic Encryption scheme, there are two significant concerns to tackle: 1.How to encode the tensor before encryption? and 2.What operations can be performed when using a particular encoding? 
The batching feature of the CKKS scheme allows a $N \times N$ matrix to be encrypted into $N$ ciphertexts, with each row or column as a ciphertext. Another possibility is to encrypt the whole tensor into a single ciphertext (\citet{jiang2018secure}). Depending on how we put our plain tensor into ciphertexts, we can perform different operations with varying complexities. The goal is to use the minimum number of ciphertexts and have a maximum depth with a minimum runtime, thus optimizing memory and computation. Seeking this ideal goal, we found out that we can use a single ciphertext to encrypt an input image and evaluate it on a convolutional neural network. This requires a pre-processing step on the client-side to encode the image as a matrix, composed of convolution windows as rows, then flatten it as a vector via a vertical scan. In TenSEAL, all these functionalities are implemented around CKKSVector.
A CKKSVector holds $N / 2$ real values, where $N$ is the polynomial modulus degree. We can perform element-wise operations with other encrypted or plain vectors (addition, subtraction, and multiplication). We have a method for computing the power of an encrypted vector (element-wise) that uses an optimal circuit, thus using a minimum multiplicative depth. Also, because we need a polynomial approximation for different activation functions, we built a method for evaluating a polynomial with the encrypted vector as a variable, making sure to use a minimal circuit. Apart from the element-wise operations, we also need matrix operations to perform machine learning tasks. We implemented a variant of the encrypted vector-plain matrix multiplication proposed by \citet{halevi} that can use multiple threads to run faster. You can check Table \ref{tab:EncryptedOps} in the appendix for a list of supported operations by the library's encrypted tensors.

\subsection{Dot Product}
The library provides a similar algorithm for dot product as in \cite{halevi}, but supports vectors of a size that is not a power of two or does not fill all the slots of a ciphertext. This limitation in the previous method was due to the right rotation that expects the final element to be the first, which was not true with regard to the cases we addressed. Our method is limited to a specific number of dot products if the vector size is not a power of two. However, this limitation is generally not reached, as the number of multiplication allowed by the scheme might be lower. We do this by replicating the input vector as many times as possible into ciphertext slots and only left ciphertext rotations during computation. Our method has the same algorithmic complexity as in \cite{halevi}. We implemented it using CKKS (\cite{cheon2017homomorphic}), for the dot product operation between an encrypted vector and a plain matrix. Thus, it can be extended to support a dot product between an encrypted matrix with a plain matrix (matrix multiplication). Figure \ref{fig:vec-matmul} in the appendix shows how to perform a dot product between an encrypted vector and a plain matrix.

\subsection{2-D Convolution}
TenSEAL also supports evaluating convolutions, with a similar implementation of how modern machine learning frameworks (e.g., PyTorch) are computing them. We applied the Image Block to Columns (im2col) (\citet{johnsoncnn}) technique, which turns a convolution layer into a single matrix multiplication operation. This technique requires encrypted matrix-plain vector multiplication, which we implemented by performing element-wise multiplication of the matrix transpose with replicated plain vector. Finally, it rotates and accumulates the result into a single vector. This operation uses only one multiplication operation and $log_2(N)$ rotations and additions, where $N$ represents the rows' number in the matrix. Section \ref{sec:2d-conv} in the appendix explains how the "image block to column" algorithm can be applied to encrypted inputs. It is important to note that the transformation happens in plain data, and the transformed input image will be encoded and encrypted in a single ciphertext. This directly implies that stacking two convolutions is not possible, as reorganizing the slots of a ciphertext is not trivial.

\section{Related work}

In recent years, several research works have made homomorphic encryption schemes practical for machine learning.  \citet{gilad2016cryptonets} implemented CryptoNets, a neural network for making inference on encrypted data using the YASHE (\cite{bos2013improved}) leveled homomorphic encryption scheme, which has efficient plain addition and multiplication algorithms, useful for unencrypted models. However, the framework requires large batches for achieving a good amortized performance, making it less practical for use cases that evaluate a single instance. \citet{boemer2019ngraph} used a similar tensor structure as CryptoNets while implementing different optimization layers. They used graph level optimizations specific for HE applications, which reduced the multiplicative depth needed for evaluating operations such as batch-norm and average pooling. In subsequent work, \citet{boemer2019ngraph2} evaluated MobileNetV2 (\cite{sandler2018mobilenetv2}) and reported empirical results on HE encrypted inputs, which was way deeper than the previously used models. \citet{jiang2018secure} used a smaller neural network and only convolution, linear layers, and the square activation function. Their framework E2DM made predictions on encrypted data using the CKKS (\cite{cheon2017homomorphic}) scheme, but compared to previous works, the model's parameters were also encrypted. \citet{juvekar2018gazelle} proposed the Gazelle framework, which mixes HE with Garbled Circuits (GC) \cite{yao1986generate}. They switched between both methods during computation, choosing the most efficient at a certain point based on the next operation. Using GC makes it possible to compute the ReLu activation function, compared to previous works (\cite{gilad2016cryptonets, jiang2018secure, boemer2019ngraph, boemer2019ngraph2}) which have mainly used polynomial functions. Even though the protocol achieves relatively fast run-time, it requires interaction between participants, resulting in high bandwidth usage. \citet{badawi2020hcnn} used a GPU-accelerated implementation of BFV(\citet{cryptoeprint:2012:144}) based on the work from \citet{bfv2018cuda}. Their HCNN of 5 layers could evaluate in 5 seconds, but could batch more than 8000 images without extra overhead.

All the works are benchmarked using the MNIST dataset (\cite{lecun2010mnist}), but with different hardware configurations. We summarize empirical results reported in each of the corresponding papers in Table \ref{table:comparison-numbers}.

\begin{table}
\centering
\begin{tabular}{ cccccc }
 \hline
 Framework & Method & Batch size & Message Size & Evaluation Time & Accuracy\\ 
 \hline \hline
 CryptoNets & HE & 8192 & 595.5 MB & 570 s & 99\%  \\
 \hline
 Gazelle & HE, MPC & 1 & 0.5 MB & 0.03 s & - \\
 \hline
 E2DM & HE & 64 & 23.93 MB & 1.69 s & 98.1\%\\
 \hline
 HCNN-GPU & HE & 8192 & - & 5.16 s & 99\%\\
 \hline
\end{tabular}
\caption{Comparing frameworks and their evaluation results on MNIST.}
\label{table:comparison-numbers}
\end{table}

\section{Evaluation}
\label{sec:evaluation}

To evaluate our library and technique, we implemented a neural network composed of: a convolutional layer (4 kernels of 7x7, with a stride of 3x3), a linear layer (input: 256, output: 64), and a final linear layer (input: 64, output: 10). We used the square activation function after every layer except for the last. The convolution was done using our image to column implementation, while the linear layers use the dot product implementation. The accuracy on the plain test-set was 97.7\% in contrast to 97.4\% for the encrypted test-set. We used the CKKSVector implementation, which uses the CKKS scheme. Knowing that we need 6 multiplications to perform the evaluation and a security level of 128-bits, we set the polynomial modulus degree to 8192, with a coefficient modulus of 206-bits, and a scale of 21-bits. The evaluation was done on Ubuntu Server 20.04 and Python 3.8, using AWS c4.2xlarge (8 vCPUs) and AWS c4.4xlarge (16 vCPUs) configurations. The measured durations are the average of 5 rounds of testing, with 10 iterations each. Table \ref{tab:mnist_benchmarks} contains a full breakdown for evaluating the neural network over encrypted images sampled from the MNIST dataset. 

\begin{table}[H]
\resizebox{\textwidth}{!}{
\begin{tabular}{llcc}
\hline
\multicolumn{1}{c}{\multirow{2}{*}{Operation}} & \multicolumn{1}{c}{\multirow{2}{*}{Description}}                                                    & \multicolumn{2}{c}{Duration(ms)}                         \\
\multicolumn{1}{c}{}                           & \multicolumn{1}{c}{}                                                                                & AWS c4.2xlarge(8 vCPUs) & AWS c4.4xlarge(16 vCPUs) \\ \hline \hline
Key generation                                 & Generate the context and the encryption keys                                                        & 940.01                     & 921.04                      \\ \hline
Input preparation                              & im2col encoding                                                                                     & 9.8                        & 9.8                         \\ \hline
Convolutional layer evaluation                 & Input $28 \times 28$, kernel $7 \times 7,$ stride 3, 4 channels & 236.9                      & 237.98                      \\ \hline
First activation(square)                       & Square $256$ input values                                                                           & 8.47                       & 8.42                        \\ \hline
FC1                                            & Fully connected layer with $256$ inputs and $64$ outputs                                            & 1084.65                    & 575.34                      \\ \hline
Second activation(square)                      & Square $64$ input values                                                                            & 4.29                       & 4.2                         \\ \hline
FC2                                            & Fully connected layer with $64$ inputs and $10$ outputs                                             & 121.36                     & 70.36                       \\ \hline
Full forward step                             & All the steps above                                                                               & 1456.29                    & 887.06    \\ \hline                  
\end{tabular}}
\caption{A complete illustration of the encrypted MNIST evaluation, with durations expressed in milliseconds. We evaluate the methods using two setups: Amazon c4.2xlarge instance (8 vCPUs, 15 GiB memory) and Amazon c4.4xlarge instance (16 vCPUs, 30 GiB memory), to underline how the library takes advantage of the available parallelism.}
\label{tab:mnist_benchmarks}
\end{table}

The results show that the library makes heavy use of the available parallelism, and it is highly competitive in terms of network communication, requiring only 427KB of communication to send the encrypted input and receive the encrypted output. At the same time, TenSEAL does not enforce a specific batch size for the inference, making it quite practical. The complete operations benchmarks are open-source, and the results are included in Section \ref{sec:complete_benchmarks} in the appendix. Table \ref{tab:unary_ops_benchmarks} shows the average performance for different arithmetic operations. Table \ref{tab:binary_ops_benchmarks} shows the average performance for matrix multiplications.

\section{Limitations and conclusion}
\label{sec:limitations}

The encryption part relies on CKKS \cite{cheon2017homomorphic}, which is known to be a leveled homomorphic encryption scheme. This means that depending on our parameter selection, there is a limit on how many multiplications we can perform on encrypted data, and this directly impacts the machine learning model we can use or its depth. Different machine learning models also use non-linear activation functions, which will need to be approximated using polynomials in the case of CKKS. A recent work \cite{chillotti2020nips} have been trying to solve this issue tightly related to machine learning by using the TFHE scheme \cite{chillotti2020tfhe}, which allows the evaluation of deeper models, as well as non-linear activation functions.

In conclusion, our results show that it can be practical to do tensorial operations using the CKKS scheme. Depending on the use case, users can choose advanced tensor operations (like slicing or broadcasting) or use more computation-communication optimized implementations. TenSEAL can accommodate both scenarios while offering a smooth transition from the traditional machine learning frameworks. Finally, we seek to extend the tensor operations catalog and to improve the overall performance even further.

\newpage

\bibliography{iclr2021_conference}
\bibliographystyle{iclr2021_conference}

\newpage

\appendix
\section{Appendix}

\subsection{Encrypted Tensor classes}
\begin{figure}[H]
   \begin{minipage}{0.48\textwidth}
     \centering
     \includegraphics[width=0.7\linewidth]{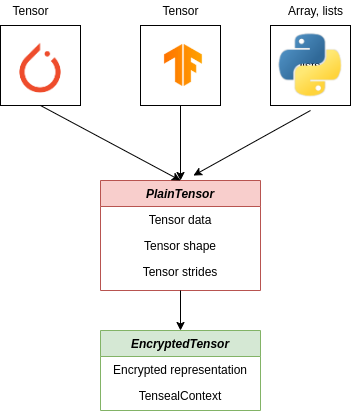}
     \caption{A high-level overview of the encrypted tensor construction. The PlainTensor wraps the tensor representation from popular frameworks, and it is used as input for the EncryptedTensor interface.}\label{fig:encryptedTensorConstruction}
   \end{minipage}\hfill
   \begin{minipage}{0.48\textwidth}
     \centering
     \includegraphics[width=0.95\linewidth]{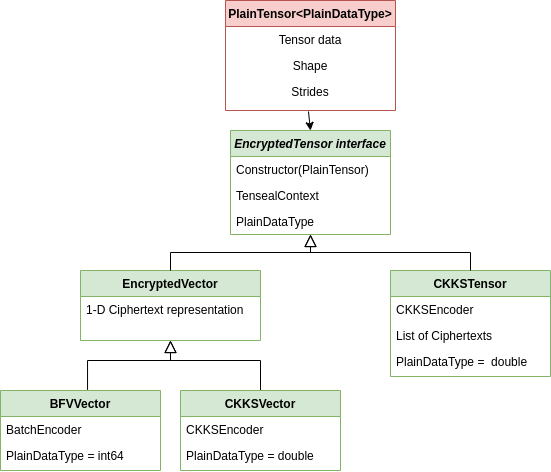}
     \caption{Encrypted tensors relation. The EncryptedTensor interface is derived into BFVVector, CKKVector, or CKKSTensor classes.}\label{fig:encryptedTensorFlow}
   \end{minipage}
\end{figure}

\subsection{Encrypted Tensor Operations}

\begin{table}[H]
\begin{tabular}{ll}
\hline
Operation                    & Description                                                   \\ \hline 
\hline
negate                       & Negate an encrypted tensor                                    \\ \hline
square                       & Compute the square of an encrypted tensor                     \\ \hline
power                        & Compute the power of an encrypted tensor                      \\ \hline
add                          & Addition between an encrypted tensor and an encrypted/plain tensor                    \\ \hline
sub                          & Subtraction between an encrypted tensor and an encrypted/plain tensor                     \\ \hline
mul                          & Multiplication between an encrypted tensor and an encrypted/plain tensor                  \\ \hline
dot\_product                 & Dot product between an encrypted tensor and an encrypted/plain tensor                     \\ \hline
polyval                      & Polynomial evaluation with an encrypted tensor as variable    \\ \hline
matmul\_plain                & Multiplication between an encrypted tensor and an encrypted/plain matrix \\ \hline
conv2d\_im2col               & Image Block to Columns                                        \\ \hline
\end{tabular}
\caption{Supported operations for encrypted tensors}
\label{tab:EncryptedOps}
\end{table}

\subsection{Dot Product}

\begin{figure}
     \centering
     \includegraphics[width=0.9\linewidth]{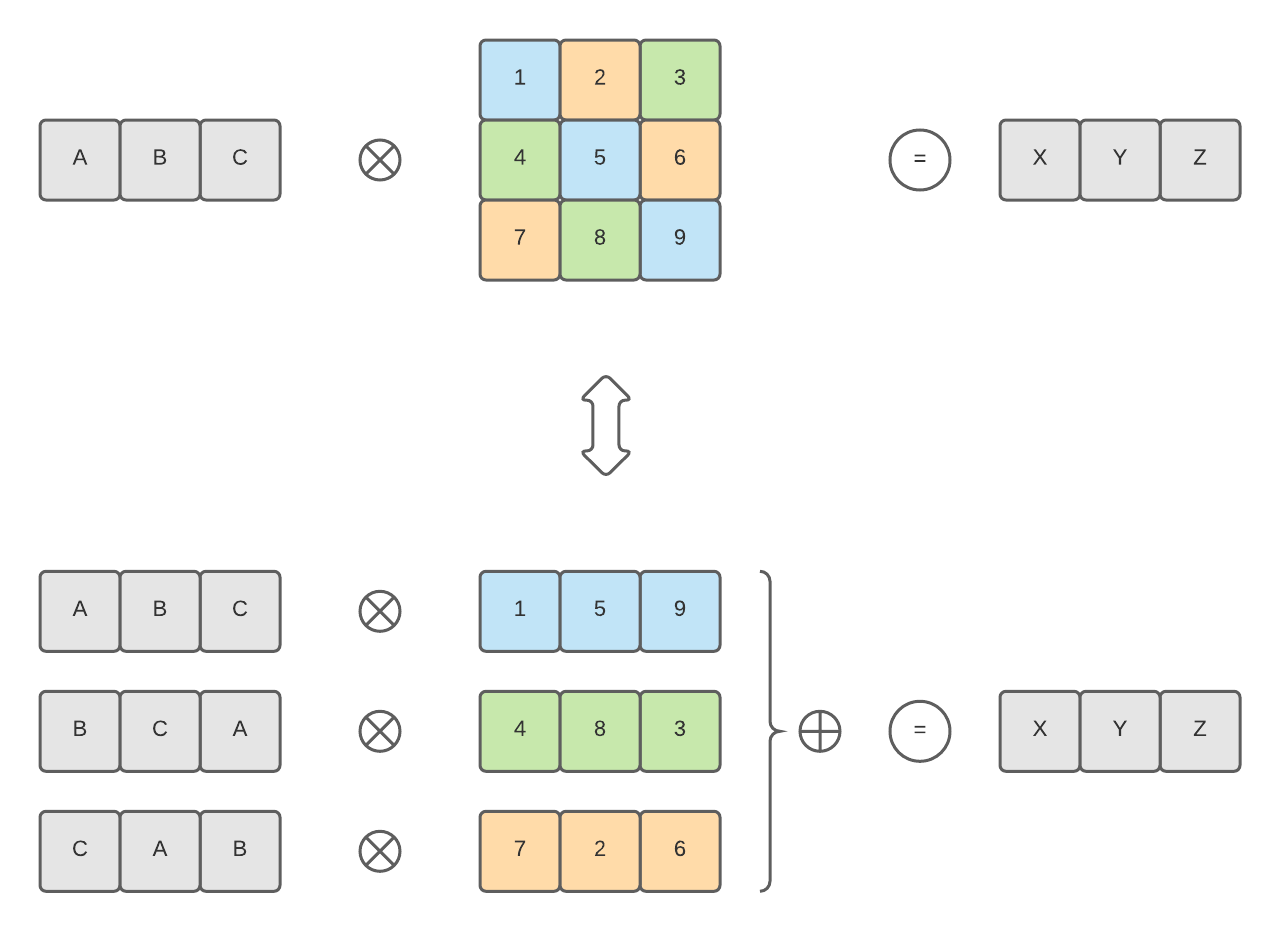}
     \caption{Vector-Matrix Multiplication}\label{fig:vec-matmul}
\end{figure}

Figure \ref{fig:vec-matmul} shows how an encrypted vector (in gray) can be multiplied with a plain matrix using the method from \cite{halevi}.

\subsection{2D Convolution}
\label{sec:2d-conv}

\begin{figure}
     \centering
     \includegraphics[width=0.9\linewidth]{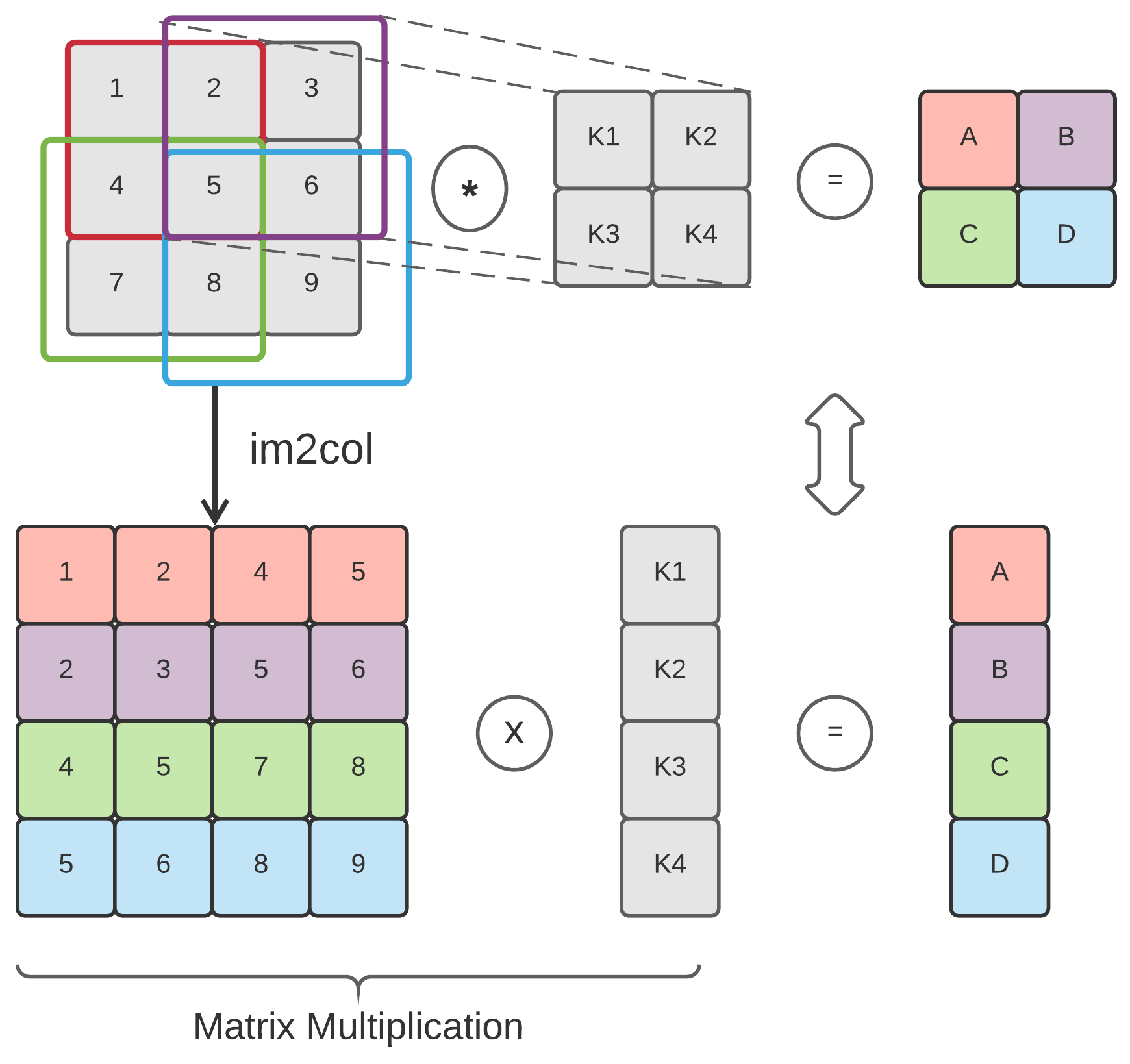}
     \caption{Image to column convolution}\label{fig:im2col_conv2d}
\end{figure}

A 2D convolution can be performed using a single matrix multiplication, instead of repeating multiplication on every window. This method is referred to as image block to column convolution, or image to column convolution. Figure \ref{fig:im2col_conv2d} shows how a convolution can be performed using this method. It first reorganizes the input matrix into rows representing convolution windows, then performs a dot product with the flattened kernel.

\begin{figure}[H]
     \centering
     \includegraphics[width=0.9\linewidth]{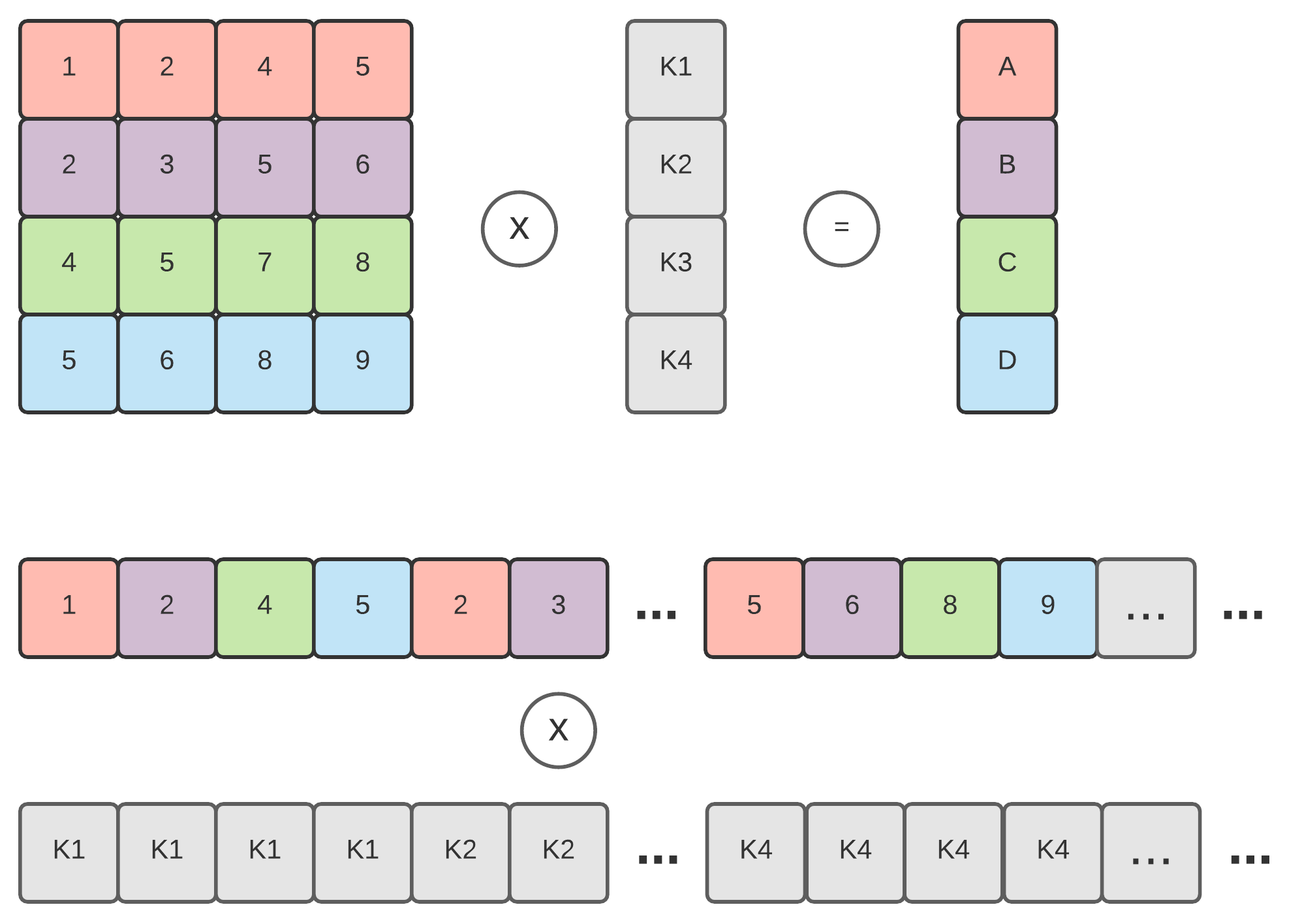}
     \caption{Image to column convolution with CKKS - step 1}\label{fig:im2col_conv2d_ckks1}
\end{figure}

\begin{figure}
     \centering
     \includegraphics[width=0.9\linewidth]{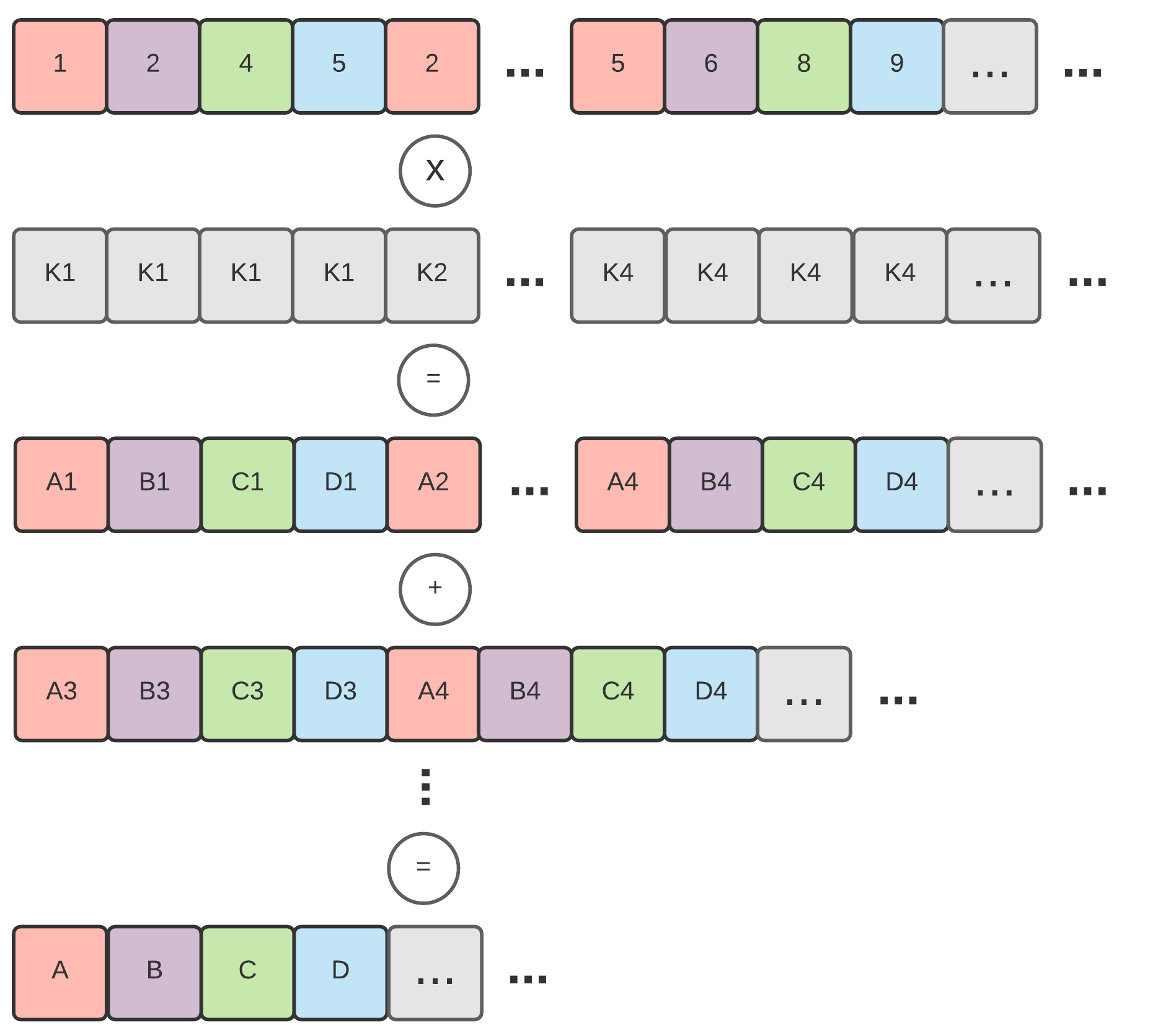}
 \caption{Image to column convolution with CKKS - step 2}\label{fig:im2col_conv2d_ckks2}
\end{figure}

Applying this technique to an encrypted matrix, which is encrypted into a single ciphertext, is not trivial, as reorganizing slots is not that simple. We will need to reorganize the matrix as a pre-processing step before encryption to be ready for convolution. The encrypted-matrix (input image) with plain-vector (kernel) can be performed with a single element-wise multiplication and a series of rotations and accumulations. Figure \ref{fig:im2col_conv2d_ckks1} and \ref{fig:im2col_conv2d_ckks2} show the steps for doing that. The first shows how the encrypted matrix (colored) is encoded and multiplied with the plain kernel. The second step is to sum different versions of the output that are rotated differently to the left.

\newpage
\subsection{Complete benchmarks}
\label{sec:complete_benchmarks}
In this section, we present the full evaluation of TenSEAL's operations, ran on the CKKSVector implementation. The benchmarks are measured on an Amazon EC2 c4.2xlarge instance, with 8 vCPUs at 2.9 GHz(Intel Xeon E5-2666 v3 Processor) and 15 GiB memory.
They are executed using Ubuntu Server 20.04 and Python 3.8. The measured durations are the average of 5 rounds of testing, with 10 iterations each.

Tables \ref{tab:unary_ops_benchmarks} and \ref{tab:binary_ops_benchmarks} show the average performance for different arithmetic operations.

\begin{table}[H]
\centering
\begin{tabular}{llllll}
\hline
\multirow{2}{*}{Operation} & \multicolumn{5}{c}{Tensor shape}                                      \\ 
                           & [256] & [1024] & [4096] & [8192] & [16384] \\ \hline \hline
negate                     & 0.07         & 0.07          & 0.07          & 0.13          & 0.26           \\ \hline
square                     & 4.29         & 4.29          & 4.29          & 8.49          & 17.16          \\ \hline
polyval                    & 10.55        & 10.46         & 10.51         & 21.32         & 42.68          \\ \hline
\end{tabular}
\caption{Duration in milliseconds for unary operations. 
The CKKS context is created for polynomial modulus $8192$ and coefficient modulus of 200-bits.
The polyval benchmark is executed for $2 X ^2 + X$ polynoms.}
\label{tab:unary_ops_benchmarks}
\end{table}

\begin{table}[H]
\centering
\begin{tabular}{llllll}
\hline
\multirow{2}{*}{Operation} & \multicolumn{5}{c}{Tensor shape}                                      \\ 
                           & [256] & [1024] & [4096] & [8192] & [16384] \\ \hline \hline
add                        & 0.08         & 0.08          & 0.08          & 0.16          & 0.31           \\ \hline
multiply                   & 4.45         & 4.34          & 4.43          & 8.84          & 17.75          \\ \hline
sub                        & 0.08         & 0.08          & 0.08          & 0.15          & 0.3            \\ \hline
dot                        & 20.15        & 23.96         & 28.11         & 55.94         & 112.36         \\ \hline
add\_plain                 & 0.8          & 0.86          & 1.07          & 2.13          & 4.19           \\ \hline
multiply\_plain            & 1.75         & 1.81          & 2.03          & 4.02          & 7.97           \\ \hline
sub\_plain                 & 0.8          & 0.86          & 1.08          & 2.14          & 4.21           \\ \hline
dot\_plain                 & 17.37        & 21.36         & 25.63         & 51.14         & 101.82         \\ \hline
\end{tabular}
\caption{Duration in milliseconds for binary operations. 
The CKKS context is created for polynomial modulus $8192$ and coefficient modulus of 200-bits. For the "\_plain" operations, the operand is a PlainTensor of the same shape. For the rest, the operand is an encrypted tensor of the same shape.}
\label{tab:binary_ops_benchmarks}
\end{table}

\end{document}

%% file: math_commands.tex
\usepackage{amsmath,amsfonts,bm}

\def\eqref#1{equation~\ref{#1}}

\def\1{\bm{1}}

\DeclareMathAlphabet{\mathsfit}{\encodingdefault}{\sfdefault}{m}{sl}
\SetMathAlphabet{\mathsfit}{bold}{\encodingdefault}{\sfdefault}{bx}{n}













%% file: main.bbl
\begin{thebibliography}{20}
\providecommand{\natexlab}[1]{#1}
\providecommand{\url}[1]{\texttt{#1}}
\expandafter\ifx\csname urlstyle\endcsname\relax
  \providecommand{\doi}[1]{doi: #1}\else
  \providecommand{\doi}{doi: \begingroup \urlstyle{rm}\Url}\fi

\bibitem[Badawi et~al.(2018)Badawi, Veeravalli, Mun, and Aung]{bfv2018cuda}
Ahmad~Al Badawi, Bharadwaj Veeravalli, Chan~Fook Mun, and Khin Mi~Mi Aung.
\newblock High-performance fv somewhat homomorphic encryption on gpus: An
  implementation using cuda.
\newblock In \emph{IACR Transactions on Cryptographic Hardware and Embedded
  Systems}, pp.\  70--95, 2018.

\bibitem[Badawi et~al.(2020)Badawi, Chao, Lin, Mun, Sim, Tan, Nan, Aung, and
  Chandrasekhar]{badawi2020hcnn}
Ahmad~Al Badawi, Jin Chao, Jie Lin, Chan~Fook Mun, Jun~Jie Sim, Benjamin
  Hong~Meng Tan, Xiao Nan, Khin Mi~Mi Aung, and Vijay~Ramaseshan Chandrasekhar.
\newblock Towards the alexnet moment for homomorphic encryption: Hcnn, the
  first homomorphic cnn on encrypted data with gpus.
\newblock In \emph{IEEE Transactions on Emerging Topics in Computing}, 2020.
\newblock \doi{10.1109/TETC.2020.3014636}.

\bibitem[Boemer et~al.(2019{\natexlab{a}})Boemer, Costache, Cammarota, and
  Wierzynski]{boemer2019ngraph2}
Fabian Boemer, Anamaria Costache, Rosario Cammarota, and Casimir Wierzynski.
\newblock ngraph-he2: A high-throughput framework for neural network inference
  on encrypted data.
\newblock In \emph{Proceedings of the 7th ACM Workshop on Encrypted Computing
  \& Applied Homomorphic Cryptography}, pp.\  45--56, 2019{\natexlab{a}}.

\bibitem[Boemer et~al.(2019{\natexlab{b}})Boemer, Lao, Cammarota, and
  Wierzynski]{boemer2019ngraph}
Fabian Boemer, Yixing Lao, Rosario Cammarota, and Casimir Wierzynski.
\newblock ngraph-he: a graph compiler for deep learning on homomorphically
  encrypted data.
\newblock In \emph{Proceedings of the 16th ACM International Conference on
  Computing Frontiers}, pp.\  3--13, 2019{\natexlab{b}}.

\bibitem[Bos et~al.(2013)Bos, Lauter, Loftus, and Naehrig]{bos2013improved}
Joppe~W Bos, Kristin Lauter, Jake Loftus, and Michael Naehrig.
\newblock Improved security for a ring-based fully homomorphic encryption
  scheme.
\newblock In \emph{IMA International Conference on Cryptography and Coding},
  pp.\  45--64. Springer, 2013.

\bibitem[Cheon et~al.(2017)Cheon, Kim, Kim, and Song]{cheon2017homomorphic}
Jung~Hee Cheon, Andrey Kim, Miran Kim, and Yongsoo Song.
\newblock Homomorphic encryption for arithmetic of approximate numbers.
\newblock In \emph{International Conference on the Theory and Application of
  Cryptology and Information Security}, pp.\  409--437. Springer, 2017.

\bibitem[Chillotti et~al.(2020{\natexlab{a}})Chillotti, Gama, Georgieva, and
  Izabach{\`e}ne]{chillotti2020tfhe}
Ilaria Chillotti, Nicolas Gama, Mariya Georgieva, and Malika Izabach{\`e}ne.
\newblock Tfhe: fast fully homomorphic encryption over the torus.
\newblock \emph{Journal of Cryptology}, 33\penalty0 (1):\penalty0 34--91,
  2020{\natexlab{a}}.

\bibitem[Chillotti et~al.(2020{\natexlab{b}})Chillotti, Joye, and
  Paillier]{chillotti2020nips}
Ilaria Chillotti, Marc Joye, and Pascal Paillier.
\newblock New challenges for fully homomorphic encryption.
\newblock \emph{Privacy-preserving Machine Learning (PPML-PriML 2020) NeurIPS
  2020 workshop}, 2020{\natexlab{b}}.

\bibitem[Fan \& Vercauteren(2012)Fan and Vercauteren]{cryptoeprint:2012:144}
Junfeng Fan and Frederik Vercauteren.
\newblock Somewhat practical fully homomorphic encryption.
\newblock Cryptology ePrint Archive, Report 2012/144, 2012.
\newblock \url{https://eprint.iacr.org/2012/144}.

\bibitem[Gilad-Bachrach et~al.(2016)Gilad-Bachrach, Dowlin, Laine, Lauter,
  Naehrig, and Wernsing]{gilad2016cryptonets}
Ran Gilad-Bachrach, Nathan Dowlin, Kim Laine, Kristin Lauter, Michael Naehrig,
  and John Wernsing.
\newblock Cryptonets: Applying neural networks to encrypted data with high
  throughput and accuracy.
\newblock In \emph{International Conference on Machine Learning}, pp.\
  201--210, 2016.

\bibitem[Halevi \& Shoup(2014)Halevi and Shoup]{halevi}
Shai Halevi and Victor Shoup.
\newblock Algorithms in helib.
\newblock pp.\  554–571, 2014.

\bibitem[Jiang et~al.(2018)Jiang, Kim, Lauter, and Song]{jiang2018secure}
Xiaoqian Jiang, Miran Kim, Kristin Lauter, and Yongsoo Song.
\newblock Secure outsourced matrix computation and application to neural
  networks.
\newblock In \emph{Proceedings of the 2018 ACM SIGSAC Conference on Computer
  and Communications Security}, pp.\  1209--1222, 2018.

\bibitem[Johnson et~al.(2016)Johnson, Li, and Karpathy]{johnsoncnn}
Justin Johnson, Fei-Fei Li, and Andrej Karpathy.
\newblock Cnns in practice. convolutional neural networks for visual
  recognitio.
\newblock 2016.

\bibitem[Juvekar et~al.(2018)Juvekar, Vaikuntanathan, and
  Chandrakasan]{juvekar2018gazelle}
Chiraag Juvekar, Vinod Vaikuntanathan, and Anantha Chandrakasan.
\newblock $\{$GAZELLE$\}$: A low latency framework for secure neural network
  inference.
\newblock In \emph{27th $\{$USENIX$\}$ Security Symposium ($\{$USENIX$\}$
  Security 18)}, pp.\  1651--1669, 2018.

\bibitem[LeCun et~al.(2010)LeCun, Cortes, and Burges]{lecun2010mnist}
Yann LeCun, Corinna Cortes, and CJ~Burges.
\newblock Mnist handwritten digit database.
\newblock \emph{ATT Labs [Online]. Available:
  http://yann.lecun.com/exdb/mnist}, 2, 2010.

\bibitem[Microsoft SEAL()]{sealcrypto}
Microsoft SEAL.
\newblock {M}icrosoft {SEAL} (release 3.6).
\newblock \url{https://github.com/Microsoft/SEAL}, November 2020.
\newblock Microsoft Research, Redmond, WA.

\bibitem[Protocol buffers()]{protobuf}
Protocol buffers.
\newblock Protocol buffers -- {Google}'s data interchange format.
\newblock URL \url{https://github.com/protocolbuffers/protobuf}.

\bibitem[Sandler et~al.(2018)Sandler, Howard, Zhu, Zhmoginov, and
  Chen]{sandler2018mobilenetv2}
Mark Sandler, Andrew Howard, Menglong Zhu, Andrey Zhmoginov, and Liang-Chieh
  Chen.
\newblock Mobilenetv2: Inverted residuals and linear bottlenecks.
\newblock In \emph{Proceedings of the IEEE conference on computer vision and
  pattern recognition}, pp.\  4510--4520, 2018.

\bibitem[TenSEAL()]{tenseal}
TenSEAL.
\newblock {TenSEAL} (release 0.3.0).
\newblock \url{https://github.com/OpenMined/TenSEAL}, February 2021.

\bibitem[Yao(1986)]{yao1986generate}
Andrew Chi-Chih Yao.
\newblock How to generate and exchange secrets.
\newblock In \emph{27th Annual Symposium on Foundations of Computer Science
  (sfcs 1986)}, pp.\  162--167. IEEE, 1986.

\end{thebibliography}
